%% file: main.tex
\documentclass[%
reprint,
superscriptaddress,
amsmath,amssymb,
aps,
]{revtex4-2}

\usepackage[T1]{fontenc} 
\usepackage[utf8]{inputenc}
\usepackage[]{graphicx}
\usepackage{blindtext}
\usepackage{lmodern}%
\usepackage{hyperref}%
\usepackage{xcolor}%
\usepackage{placeins}
\usepackage{bm}
\usepackage{tabularx}
\usepackage{booktabs}
\usepackage[normalem]{ulem}

\include{incs}

\newcommand{\vg}{\ensuremath{V_{\rm G}}}

\newcommand{\dqmp}{Department of Quantum Matter Physics, University of Geneva, 24 Quai Ernest Ansermet, CH-1211 Geneva, Switzerland}
\newcommand{\gap}{Group of Applied Physics, University of Geneva, 24 Quai Ernest Ansermet, CH-1211 Geneva, Switzerland}
\newcommand{\moe}{MOE Key Laboratory for Non-equilibrium Synthesis and Modulation of Condensed Matter, Shaanxi Province Key Laboratory of Advanced Materials and Mesoscopic Physics, School of Physics, Xi'an Jiaotong University,  Xi’an 710049, China.}
\newcommand{\KW}{Research Center for Functional Materials, NIMS, 1-1 Namiki, Tsukuba 305-0044, Japan}
\newcommand{\TT}{International Center for Materials Nanoarchitectonics, NIMS, 1-1 Namiki, Tsukuba 305-0044, Japan} 
\newcommand{\ngi}{National Graphene Institute, University of Manchester, Manchester M13 9PL, UK}
\newcommand{\manchester}{School of Physics \& Astronomy, University of Manchester, Manchester M13 9PL, UK}
\newcommand{\royce}{Henry Royce Institute for Advanced Materials, Manchester M13 9PL, UK}

\definecolor{AAcolor}{rgb}{0.7,0.1,0.4}

\usepackage{newfloat}

\definecolor{linkcol}{rgb}{0,0,0.4}
\definecolor{citecol}{rgb}{0.5,0,0}
\hypersetup{
	bookmarksopen=true,
	pdftitle="Band gap opening in Bilayer Graphene-CrCl3/CrBr3/CrI3 van der Waals Interfaces",
	pdfauthor="Tenasini et al",
	pdfsubject="Band gap opening in Bilayer Graphene-CrCl3/CrBr3/CrI3 van der Waals Interfaces", 
	pdfstartview={FitH}, 
	pdfmenubar=true, 
	pdfhighlight=/O, 
	colorlinks=true, 
	pdfpagemode=UseNone, 
	pdfpagelayout=SinglePage, 
	pdffitwindow=true, 
	linkcolor=linkcol, 
	citecolor=linkcol, 
	urlcolor=blue
}

\begin{document}
	
	\title{Band Gap Opening in Bilayer Graphene-CrCl$_3$/CrBr$_3$/CrI$_3$ van der Waals Interfaces}
	
	\author{Giulia Tenasini}
	\email{giulia.tenasini@unige.ch} 
	\affiliation{\dqmp} \affiliation{\gap} 
	
	\author{David Soler-Delgado} 
	\affiliation{\dqmp} \affiliation{\gap}
	
	\author{Zhe Wang}\affiliation{\dqmp} \affiliation{\moe}
	
	\author{Fengrui Yao}\affiliation{\dqmp} \affiliation{\gap}     
	
	\author{Dumitru Dumcenco} \affiliation{\dqmp}      
	
	\author{Enrico Giannini} \affiliation{\dqmp}        
	
	\author{Kenji Watanabe} \affiliation{\KW}
	
	\author{Takashi Taniguchi} \affiliation{\TT}
	
	\author{Christian Moulsdale} \affiliation{\ngi}  \affiliation{\manchester} 
	
	\author{Aitor Garcia-Ruiz} \affiliation{\ngi} \affiliation{\manchester}         
	
	
	\author{Vladimir I. Fal’ko} \affiliation{\ngi} \affiliation{\manchester} \affiliation{\royce}           
	
	\author{Ignacio Gutiérrez-Lezama}\affiliation{\dqmp} \affiliation{\gap} 
	
	\author{Alberto F. Morpurgo} \email{alberto.morpurgo@unige.ch}\affiliation{\dqmp} \affiliation{\gap}

	
	\date{\today}
	
	

	\begin{abstract} 
   We report experimental investigations of transport through bilayer graphene (BLG)/chromium trihalide (CrX$_3$; X=Cl, Br, I) van der Waals interfaces. In all cases, a large charge transfer from BLG to CrX$_3$ takes place (reaching densities in excess of $10^{13}$~cm$^{-2}$), and generates an electric field perpendicular to the interface that opens a band gap  in BLG. We determine the gap from the activation energy of the conductivity and find excellent agreement with   the latest theory accounting for the contribution of the $\sigma$ bands to the BLG dielectric susceptibility. We further show that for BLG/CrCl$_3$ and BLG/CrBr$_3$ the band gap can be extracted from the gate voltage dependence of the low-temperature conductivity, and use this finding to refine the gap dependence on the magnetic field. Our results allow a quantitative comparison of the electronic properties of BLG with theoretical predictions and indicate that electrons occupying the CrX$_3$ conduction band are correlated. 
    \end{abstract}

	\maketitle

Van der Waals (vdW) interfaces provide a vast playground for creating new systems with engineered electronic properties, by stacking suitably chosen atomically thin crystals (or 2D materials) on top of each other. Examples include hexagonal Boron Nitride (hBN) encapsulation of graphene \cite{Varlet2014, Palau2018, finney_tunable_2019}, proximity induced  spin-orbit coupling in graphene on semiconducting transition metal dichalcogenide substrates \cite{Avsar2014,Wang2015,wang_origin_2016,Gmitra2017,ghiasi_large_2017,Khoo2017,benitez_strongly_2018}, or the creation of so-called $\Gamma-\Gamma$ interfaces  \cite{Terry_2018,Ubrig2020}. Recently, the discovery of 2D magnets and their use in  vdW heterostructures has further broadened the scope of phenomena that can be explored \cite{zhong_van_2017, gibertini_magnetic_2019, gong_two-dimensional_2019,mak_probing_2019,huang_emergent_2020, Kurebayashi2022}. In these systems, the wave functions of electrons in the non-magnetic material  extend into the magnetic one and experience  some of the magnetic interaction,   enabling magnetism to be proximity induced in graphene or other 2D materials \cite{Qiao2014, Leutenantsmeyer2016, zhong_van_2017,  Tang2017, Seyler2018, Cardoso2018, Farooq2019, Tang2020, Wu2021, Vila2021, Ghiasi2021}. However, deterministically controlling  magnetism by proximity and predicting what aspect of magnetism can be induced into non-magnetic materials  are challenges that remain to be solved because many different phenomena – such as strain, hybridization, charge transfer, and more – occur simultaneously at van der Waals interfaces \cite{Ugeda2014, gong_two-dimensional_2019, PhysRevApplied.12.024031, Zhou2019}, influencing the interfacial electronic properties. In particular, electrostatic effects often dominate the behavior of heterostructures formed by low-charge-density systems, such as 2D semiconductors and semimetals. As a result, significant charge transfer can occur and lead to new phenomena mediated by changes in electron concentration or orbital occupation \cite{Jiang2018}. Indeed, recent work reported spin-dependent interlayer charge transfer in magnetic vdW heterostructures \cite{Zhong2020, Lyons2020, mashhadi_spin-split_2019} and concluded that its detailed analysis is of key importance for improving the control of interfacial properties. \\

\begin{figure*}[ht!]
    \centering
	\includegraphics[width=\textwidth]{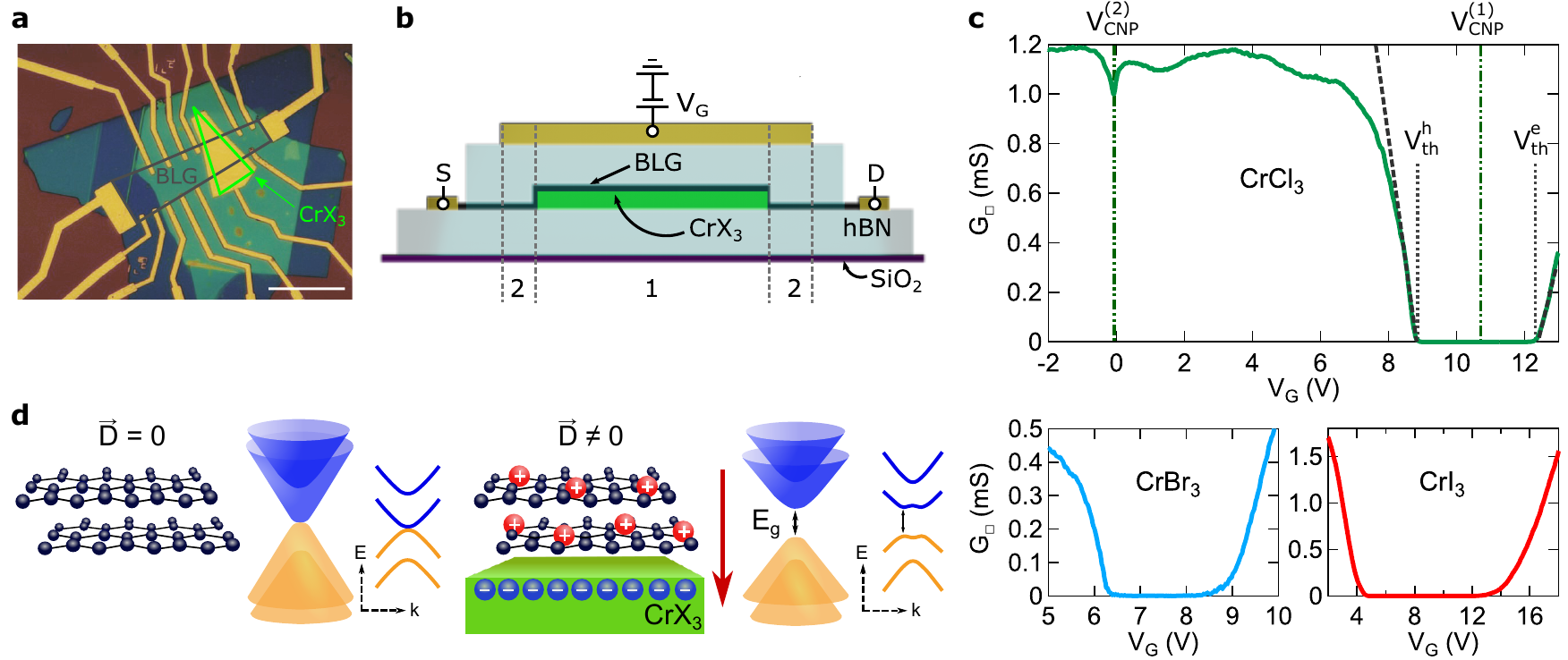}
	\caption{\textbf{Band gap opening in BLG/CrX$_3$ interfaces.} \textbf{(a)} Optical micrograph and \textbf{(b)} schematics of a representative device (the scale bar in (a) is 20~$\mu$m), based on a  BLG/$\rm{CrX_3}$ heterostructure  encapsulated in hBN. A metallic gate electrode is deposited onto the top hBN layer, and is coupled to two distinct regions: a central part formed by the BLG/$\rm{CrX_3}$ interface (region 1) and two adjacent parts where BLG is in contact only with hBN (region 2). Transport is measured using metallic source (S) and drain contacts (D) and probes the two regions connected in series. \textbf{(c)} Square conductance $G_{\square}$  as a function of gate voltage \vg\ measured in a heterostructure of BLG-on-CrCl$_3$ at 250 mK (green curve, top). Two characteristic features are visible in the transfer curve: a small conductance dip close to \vg~=~0~V corresponding to the CNP of graphene in region 2 ($V_{CNP}^{(2)}$) and a pronounced suppression at large \vg\ ($V_{CNP}^{(1)}$) that originates from gating BLG-on-CrCl$_3$ (i.e., region 1) to charge neutrality. The black dashed lines represent linear extrapolations to extract the threshold voltages for holes $V^h_{th}$  and electrons $V^e_{th}$. Virtually identical behaviour is observed in heterostructures of BLG and CrBr$_3$ (light-blue, bottom) and  BLG-on-CrI$_3$ (red, bottom). \textbf{(d)} Schematics of the BLG band structure in the absence (left) and presence (right) of a perpendicular displacement field $\vec{D}$, showing that at finite field a gap $E_g$ is present at charge neutrality. As visible on the right side of the panel, the displacement field is generated by the large transfer of electrons from BLG to CrX$_3$ occurring at the vdW interface.}
	\label{fig:1}
\end{figure*}%

Here, we report the systematic behavior of vdW interfaces formed by bilayer graphene (BLG) \cite{Novoselov2006, McCann2006} and chromium trihalide crystals (CrX$_3$; X=Cl, Br, I) \cite{Wang_2011, Zhang2015, McGuire2015, Liu2016, Huang2017, Wang2018, Wang2019, PhysRevB.104.155109}. All systems exhibit a large transfer of electrons from graphene to the magnetic material, reaching values in excess of $10^{13}$ cm$^{-2}$, producing a large electric field perpendicular to the interface and a gap in BLG. When gating the BLG at the charge neutrality point (CNP), the gap induces a low-temperature suppression of the conductance of  four orders of magnitude or more, exhibiting a sharp onset as a function of gate voltage. We determine the size of the band gap by analyzing the temperature dependence of the transfer curves (i.e., conductance-vs-gate voltage), and find excellent agreement with the results of the latest ab initio calculations of the electrostatically induced gap in BLG, which include dielectric screening due to the polarizability of the $\sigma$ bonds in the graphene lattice \cite{Slizovskiy2021}. We also find that the gap in BLG can be determined by looking exclusively at the low-temperature gate voltage dependence of the conductance, a result that provides information about the nature of the electronic states in the CrX$_3$, and that enables the quantitative determination of the dependence of the BLG band gap on the applied magnetic field. \\

Figure \ref{fig:1}a shows an optical microscope image of a representative device employed for our transport studies. A BLG with an elongated rectangular shape is placed on top of an exfoliated CrX$_3$ thin crystal, using a by-now conventional dry transfer method \cite{zomer_transfer_2011}. The process is carried out in the controlled environment of a glove box and the interface is encapsulated in hBN to prevent degradation upon exposure to air. Metal contacts to BLG are patterned using standard micro-fabrication techniques, and a gate electrode is deposited onto the top hBN, enabling the charge density in the BLG layer to be tuned. The device schematics in Figure~\ref{fig:1}b highlights how the gate is coupled to two distinct regions: a central part formed by the BLG/CrX$_3$ interface (which we refer to as region 1) and two parts on the sides, where BLG is in contact only with hBN (which we refer to as region 2) that effectively act as contacts to the gapped part of the structure. This device feature is important to understand some aspects of the measurements that we present later. \\

The gate voltage dependence of the square conductance $G_{\square}$ of a device with BLG-on-CrCl$_3$, shown in Figure~\ref{fig:1}c (green curve), reveals two characteristic features that originate from regions 1 and 2. The small conductance dip close to \vg~=~0~V is the manifestation of the charge neutrality point (CNP) of BLG-on-hBN (region 2), and the pronounced suppression (four orders of magnitude) at large \vg\ originates from having gated BLG-on-CrCl$_3$ (region 1) to charge neutrality. The shift of the BLG CNP toward high, positive gate voltages indicates that a large number of electrons are transferred from BLG to the CrCl$_3$ crystal ($10^{13}$~cm$^{-2}$). Analogous behavior and a large hole doping in BLG (see Section S1 of the Supporting Information for transfer curves in a broader range of gate voltages) are also observed in heterostructures formed by BLG and CrBr$_3$ (light-blue curve) and in BLG-on-CrI$_3$ (red curve; in agreement with the earlier observations \cite{Jiang2018, Jiang2019}). \\

\begin{figure*} 
	\centering
	\includegraphics[width=\textwidth]{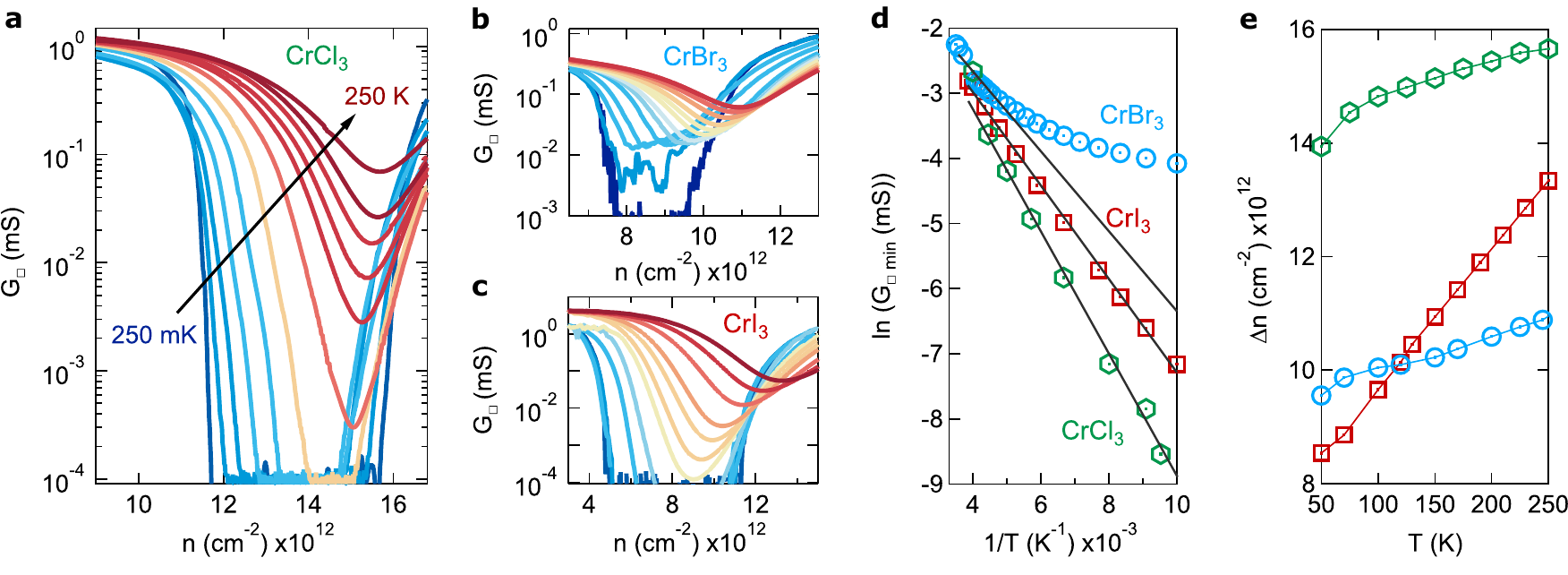}%
	\caption{\textbf{Temperature evolution of the square conductance in BLG/CrX$_3$  heterostructures.} Square conductance $G_{\square}$ as function of charge density $n$ measured at different temperatures between 250 mK (blue) and 250 K (red) for \textbf{(a)} BLG-on-CrCl$_3$,\textbf{(b)} CrBr$_3$  and \textbf{(c)} CrI$_3$. \textbf{(d)} Arrhenius plot of the minimum square conductance measured for the three different heterostructures (CrCl$_3$ green hexagons,  CrI$_3$ red squares, CrBr$_3$ light-blue circles). Activation energies are obtained by fitting the linear part in the high-temperature range (grey lines). \textbf{(e)} Temperature dependence of the charge $\Delta n$ transferred from BLG to CrX$_3$  for the investigated interfaces (the different colors and symbols represent data measured on different interfaces, as indicated in (d)).
    }
    \label{fig:2}
\end{figure*}%
	
\begin{figure}
	\centering
	\includegraphics[width=0.485\textwidth]{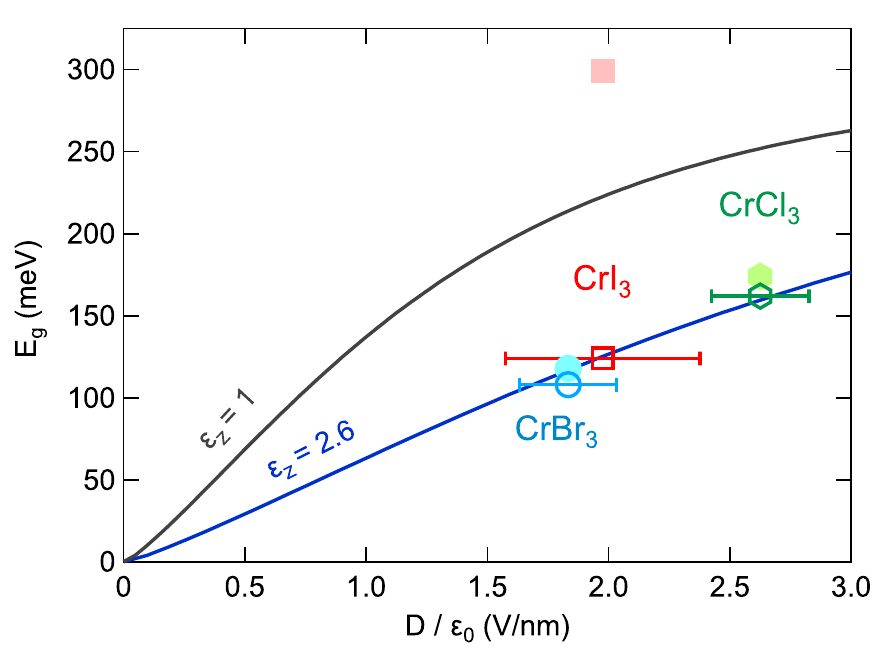}%
	\caption{\textbf{Electric field dependence of the band gap in BLG/CrX$_3$ interfaces.} The continuous lines represent the band gap as a function of displacement field $D$ predicted by ab initio calculations, considering or ignoring the contribution to the dielectric susceptibility $\varepsilon_z$ due to the electrons that occupy the $\sigma$ band of  BLG \cite{Slizovskiy2021}. The empty symbols represent the experimental data obtained from the temperature dependence of the conductance measured in our devices. It is apparent that the experimental data are in excellent agreement with  theoretical prediction for $\varepsilon_z=2.6$. The error bars for the displacement field ($D = e \Delta n$, see main text) correspond to the variation of charge transferred from BLG to CrX$_3$ (and consequently of $D$) as temperature is varied. Filled symbols indicate the experimental values of $E_g$ extracted from the threshold voltages of low-temperature transfer curves using Equation (1). For CrCl$_3$ and CrBr$_3$ the agreement with the gap values obtained from the temperature-dependent measurements is excellent.  }
	\label{fig:3}
\end{figure}

The observed charge transfer generates a strong electric field perpendicular to the interface that causes the opening of a band gap in BLG \cite{PhysRevB.74.161403, McCann2006, Ohta2006, Castro2007, Oostinga2008, Kuzmenko2009, Zhang2009} (see the schematic band diagrams in Figure~\ref{fig:1}d). As a result, when the chemical potential in BLG is shifted to charge neutrality by applying a suitable gate voltage, this leads to a robust insulating state. The sharp onset of this insulating state as a function of \vg\ implies the absence of large electrostatic potential fluctuations at the BLG/CrX$_3$ interfaces, indicating that charge transfer from CrX$_3$ to BLG is rather homogeneous. To compare quantitatively the experimental observations made in heterostructures based on the different CrX$_3$, we convert the applied gate voltage to the corresponding accumulated charge density $n$, as $n=\frac{\varepsilon \,\varepsilon_{0}}{\mathrm{t}} \frac{V_{\mathrm{G}}-V_{\mathrm{CNP}}^{(2) }}{e}$ ($\varepsilon$ and $t$ are the relative dielectric constant and thickness of the hBN layer, and $V_{CNP}^{(2)}$ is the gate voltage corresponding to the CNP in region 2, see  Figure~\ref{fig:1}c). The concentration of electrons transferred from BLG in CrX$_3$ is then given by $\Delta n = \frac{\varepsilon \, \varepsilon_0}{t}\frac{V_{CNP}^{(1)} - V_{CNP}^{(2)}}{e}$ (where $V_{CNP}^{(1)}$ is the  gate voltage corresponding to the  CNP of BLG-on-CrX$_3$, i.e. in region 1; see  Figure~\ref{fig:1}c again), and  is directly proportional to the displacement field present at the BLG/CrX$_3$ interfaces, $D = e \Delta n$ (we use this relation  to calculate the values of the displacement field in Figure~\ref{fig:3}). \\

We determine the size of the band gap from the temperature ($T$) evolution of the transfer curves, plotted in Figure~\ref{fig:2}a-c for CrCl$_3$, \CrBrThree, and CrI$_3$, respectively. Data are shown for selected values of $T$ between 250 mK (blue curve) and 250 K (red). The activation energies for the three heterostructures are extracted by looking at the minimum square conductance $G_{\square}^{min}$  (corresponding to $G_{\square}$ at the CNP of BLG/CrX$_3$), by fitting the linear part of the Arrhenius plot in the high-temperature range, where the charge carriers are dominated by  a thermally activated behavior  (see Figure~\ref{fig:2}d). The larger activation energy $E_a$ is observed in BLG-on-CrCl$_3$ (green hexagons) where the gap ($E_g = 2 E_a$) is estimated to be  162~meV; band gaps of 124~meV and 108~meV are found for BLG-on-CrI$_3$ (red squares) and on CrBr$_3$ (light-blue circles), respectively. \\

The same measurements show that the position of the CNP, i.e., the density of charge transferred from BLG to CrX$_3$, is temperature-dependent (as summarized in Figure~\ref{fig:2}e), implying that the perpendicular electric field responsible for the opening of the band is not constant as $T$ is varied. For BLG-on-CrCl$_3$ (green) and on CrBr$_3$ (light-blue) the position of CNP changes by less than 10\% throughout the full range investigated, and by significantly less over the range used to determine the size of the band gap, so that the effect can be disregarded.  For CrI$_3$ (red) the change is larger, corresponding to a more sizable indetermination for the electric field value responsible for the opening of the gap in BLG. The precise microscopic origin of the $T$ dependence of CNP in BLG-on-CrI$_3$ is currently not understood, and is likely determined by the electronic properties of CrX$_3$, which are materials with very narrow bands, whose behavior deviates from that of conventional semiconductors (see also the below discussion on the determination of the gap from the gate voltage dependence of the transfer curves). \\

The band gap dependence on the electric field for the three different BLG/CrX$_3$ interfaces is compared to the calculated gap in Figure~\ref{fig:3}. The  electric field dependence of the gap predicted for $\varepsilon_z = 2.6$ –corresponding to the theoretically expected dielectric susceptibility when accounting for the polarizability of the $\sigma$ bands– is represented by the blue line. For comparison we also show calculations with  $\varepsilon_z = 1$ (grey line) as in \cite{McCann2006}. The empty symbols of different colors (CrCl$_3$, green hexagon;  CrI$_3$, red square; CrBr$_3$, light-blue circle) represent our experimental data, and the error bar denotes the indetermination on the electric field due to the temperature dependence of the charge transferred from BLG to CrX$_3$, as just discussed above. These data agree perfectly with theory that considers $\varepsilon_z = 2.6$ using the method proposed in \cite{Slizovskiy2021}, and deviate very significantly from the $\varepsilon_z = 1$ curve. This result should be underscored, because experimental values for the band gap reported in the early days of research on graphene were larger, and it was argued  that quantitative agreement was obtained for $\varepsilon_z = 1$ (the deviation  likely originated from a insufficiently sharp dependence of the conductance on \vg\, due to the lower quality of  the BLG-on-SiO$_2$ devices used in earlier experiments \cite{Zhang2009}).\\

Having determined the band gap from the analysis of the temperature dependence of the square conductance, we now present an alternative way that relies exclusively on the analysis of the low-temperature G$_\square$-vs-\vg\ curves. From analyzing our data, we find that for BLG on both CrCl$_3$ and CrBr$_3$, the gap is quantitatively given by
\begin{equation}
E_g=\frac{C\left(V_{t h}^{e}-V_{t h}^{h}\right)}{e \ \rho_{\mathrm{BLG}}}
\end{equation}
Here, $V^e_{th}$ and $V^h_{th}$ are the threshold voltages for electron and hole conductance (obtained by extrapolating to zero the conductance measured as a function of gate voltage, as illustrated by the dashed lines in Figure~\ref{fig:1}c) and $\rho_{\mathrm{BLG}}=2m^*/\pi h^2$ is the density of states in gapless BLG contacts (i.e., region 2 in Figure~\ref{fig:1}b), next to the gapped region where BLG is on the CrX$_3$ layer (region 1). Equation (1) gives the gap values represented with filled symbols in Figure~\ref{fig:3}, in perfect agreement with the values obtained from the $T$ dependence of the minimum conductance for both BLG-on-CrCl$_3$ and BLG-on-CrBr$_3$; for BLG-on-CrI$_3$, instead, Eq. (1) gives a value that deviates by nearly a factor of 2 from the correct one. We discuss below the origin of Eq. (1), the condition for its validity, and why it fails to give the correct value of the gap for CrI$_3$. \\

\begin{figure}
	\centering
	\includegraphics[width=0.483\textwidth]{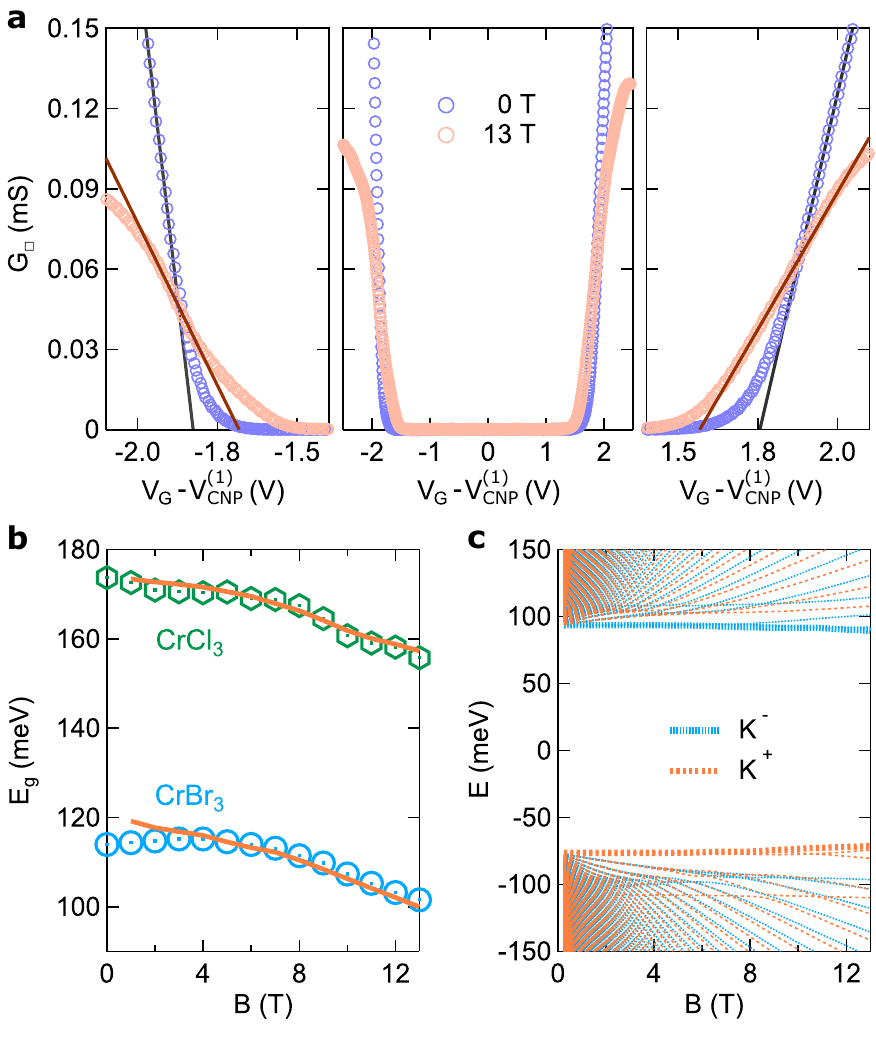}
	\caption{\textbf{Magnetic field dependence of the BLG band gap.} \textbf{(a)}, Square conductance $G_{\square}$  as a function of gate voltage \vg\ shifted with respect to the value of CNP $V_{CNP}^{(1)}$, measured in a BLG/CrCl$_3$ heterostructure at 0 T and with an applied magnetic field of 13 T. The left and the right panels zoom in on the onset of conduction for holes and electrons: the corresponding threshold voltages $V_{th}^h$ and $V_{th}^e$ shift upon increasing the magnetic field, resulting in a decrease in ($V_{th}^e$-$V_{th}^h$) and therefore in a decrease in the band gap extracted using Eq. (1). \textbf{(b)}, Magnetic field dependence of the energy gap for BLG-on-CrCl$_3$ (green empty hexagons) and for BLG-on-CrBr$_3$ (light-blue empty circles); the size of the symbols corresponds to the experimental uncertainty associated with the error in the determination of the threshold voltages. The continuous orange lines represent the calculated band gap considering appropriate screened interlayer asymmetry potentials and including a screening potential for non-zero magnetic fields, as predicted by theory to calculate the Landau level spectrum. The experimental data are in excellent agreement with the theoretical predictions.  \textbf{(c)}, Landau levels calculated for a screened interlayer asymmetry potential of $|\Delta|=215$~meV resulting in an experimentally observed gap of $E_g=170$~meV at zero applied magnetic field. The dependence of the gap on the magnetic field is determined by the difference in  the energies of the lowest Landau level in the conduction and valence bands.}
	\label{fig:4}
\end{figure}%
	
The possibility to use Eq. (1) to extract the band gap of BLG enables the detailed dependence of the gap on the magnetic field, not yet addressed in previous studies, to be probed in an experimentally straightforward way.  To this end, it suffices to measure the conductance as a function of gate voltage for different values of the magnetic field, as shown in Figure~\ref{fig:4}a for a BLG-on-CrCl$_3$ device. The left and right panels zoom in on the onset of threshold for both electron and hole conduction, which shift upon increasing the applied magnetic field, resulting in a decrease in  ($V_{th}^e$-$V_{th}^h$), and therefore a decrease in the BLG band gap. The full dependence of the gap on $B$ for BLG-on-CrCl$_3$ and for BLG-on-CrBr$_3$ (represented by the green hexagons and light-blue circles, respectively) is compared to the theoretically calculated dependence (orange line) in Figure~\ref{fig:4}b. Theory predicts that the gap decreases as a result of the formation of Landau levels \cite{Varlet2014}, which causes the top of the valence band to increase in energy and the bottom of the conduction band to decrease (see Figure~\ref{fig:4}c). This is a counter-intuitive behavior that contrasts what would be expected for electrons in a conventional two-dimensional electron gas (i.e., electrons described by a scalar wave function, for which the formation of Landau levels would lead to an increase in the gap at finite $B$). At a quantitative level, a change in the gap between 10\% and 15\% is expected as $B$ increases up to 13~T, which results in perfect agreement with experiments without the need to introduce any free fitting parameters. To confirm the soundness of this result and exclude significant contributions of other mechanisms to the observed magnetic field dependence of the band gap, we also considered whether the charge transferred from BLG to CrX$_3$ depends on the magnetic field. This is important because a change in charge transfer would lead to a corresponding change in the perpendicular electric field and thus in the size of the band gap. Nevertheless, the analysis of the magnetic field dependence of the CNP (Section S2 of the Supporting Information) shows that, even if charge transfer slightly decreases at high magnetic fields, the quantitative effect on the gap is very small, close to the sensitivity of the experiment, and negligible in a first approximation. \\

Such excellent quantitative agreement shows the usefulness of Eq. (1) and confirms its validity. To understand heuristically the origin of Eq. (1) we look at how the electrostatic and electrochemical potentials vary in the different regions of our devices (see Figure~\ref{fig:1}b), i.e., in region 1 where BLG is in contact with the CrX$_3$ layer and in region 2 where BLG is on hBN. A change $\Delta V_{\rm{G}}$ in applied gate voltage causes a variation in the electrochemical and electrostatic potentials $\Delta \mu$ and $\Delta \phi$ in both regions, with the two quantities related by $\Delta \mu=e \Delta \phi+\Delta E_F$  ($\Delta E_F$ is the change in Fermi energy induced by the variation in the density of accumulated electrons in BLG, i.e., $\Delta E_F = C\Delta V_{\rm{G}} /e\rho_{\mathrm{BLG}}$). Since the entire structure is at equilibrium for all gate voltages, the change in electrochemical potential is uniform, such that $\Delta \mu^{(1)}=\Delta \mu^{(2)}$, or $e\Delta \phi^{(1)}+\Delta E_F^{(1)}=e \Delta \phi^{(2)}+\Delta E_{F}^{(2)}$. Whenever the electrochemical potential in region 1 is inside the gap of BLG –and at sufficiently low temperature- $ \Delta E_{\mathrm{F}}{ }^{(1)}=0$, because no states are available to add charge. Under these conditions, therefore, a variation in gate voltage only changes the electrostatic potential in region 1, so that we have $e \Delta \phi^{(2)}+\Delta E_{F}^{(2)}=e \Delta \phi^{(1)}$. As $\Delta E_{\mathrm{F}}^{(2)}=C \Delta V_{\mathrm{G}} / e\rho_{\mathrm{BLG}}$, we obtain $e \Delta \phi^{(1)}-e \Delta \phi^{(2)}=C \Delta V_{\mathrm{G}} / e\rho_{\mathrm{BLG}}$, a relation that determines the relative band alignment between region 1 and 2. While we sweep the gate voltage from $V_{\rm{G}} = V_{th}^h$ to $V_{\rm{G}} = V_{th}^e$, this relation always holds, because throughout this \vg\ interval the electrochemical potential in region 1 is inside the gap. Since at $V_{\rm{G}} = V_{th}^h$ the electrochemical potential in region 2 is aligned with the valence band edge in region 1 and at $V_{\rm{G}} = V_{th}^e$ the electrochemical potential in region 2 is aligned with the conduction band edge in region 1, we obtain $E_g=e \Delta \phi^{(1)}-e \Delta \phi^{(2)}$, and Eq. (1) then follows directly using that $e \Delta \phi^{(1)}-e \Delta \phi^{(2)}=C \Delta V_{\mathrm{G}} / e\rho_{\mathrm{BLG}}$. We interpret this result by saying that a change in \vg\ lowers the bands of BLG in region 1 and in region 2 (which effectively forms the source and drain contacts to the transistor channel), but changes $E_F$ only in region 2 (because only region 2 is gapless), and it is this change in Fermi energy that shifts the electrochemical potential from the valence to the conduction band edge. \\

The argument above relies on the assumption that charge transferred from BLG to the underlying CrX$_3$ layer is fixed: at low temperature, a change in \vg\ does not change the charge accumulated in the CrX$_3$ layer. This is not what would happen if electrons in CrX$_3$ behaved as independent, non-interacting particles, i.e., if CrX$_3$ could be described as a conventional semiconductor. We attribute  this behavior to the very narrow bands of CrX$_3$ –electrons added to CrX$_3$ are virtually localized on the Cr orbitals– which make electrons hosted in these materials strongly correlated, because the strength of their Coulomb interaction is larger than the bandwidth. As a result, at low temperature, the electrons transferred from BLG to the surface of CrX$_3$ create an energetically stable correlated state (we imagine a spatially ordered distribution of electrons localized on Cr atoms that minimizes energy), which has an energy gap for adding or removing electrons. This assumption appears to be fully consistent with the behavior observed in BLG interfaces with CrCl$_3$ and CrBr$_3$, for which Eq. (1) works perfectly, but not for BLG-on-CrI$_3$, for which Eq. (1) gives a factor of 2 deviation as compared to the actual gap. The reason for this difference between the different CrX$_3$ compounds likely originates from the fact that the charge transferred from BLG to CrI$_3$ does vary as \vg\ is varied (possibly because the width of the conduction band of CrI$_3$ is somewhat larger than that of CrCl$_3$ and CrBr$_3$), an observation that seems consistent with the pronounced temperature dependence of charge transfer from BLG to CrI$_3$ (see Figure~\ref{fig:2}e). This conclusion underscores the unconventional nature of the semiconducting properties of chromium trihalides, which calls for more detailed future investigations, and the fact that the study of transport through BLG/CrX$_3$ interfaces allows differences in the electronic properties of the different members of this family to be evidenced. \\

In summary, we have performed a systematic analysis of different phenomena  determining the  transport properties of vdW interfaces based on BLG and chromium trihalide crystals (CrCl$_3$, CrBr$_3$ and CrI$_3$). In all cases, a very large charge transfer from graphene to CrX$_3$ is found to occur, which  causes the opening of a band gap in BLG. A detailed comparison shows that the values of the gap determined experimentally are  in  excellent agreement with the latest ab initio calculations, which  include the effect of the polarizability of the $\sigma$ bands in the graphene honeycomb lattice. We furthermore show that it is possible to determine the band gap quantitatively by looking exclusively at the low-temperature gate voltage dependence of the conductivity, a finding that we exploit to determine how the band gap depends on the applied magnetic field. Besides providing indications as to the correlated nature of electrons transferred onto the very narrow conduction band of CrX$_3$, our  work establishes a remarkable  quantitative agreement between different electronic properties of BLG and corresponding  theoretical predictions. 

\vspace{0.1cm}

\section*{Supporting Information}
The Supporting Information is available free of charge on the ACS publications website at \href{https://pubs.acs.org/doi/10.1021/acs.nanolett.2c02369}{https://pubs.acs.org/doi/10.1021/acs.nanolett.2c02369}.
\begin{itemize}
    \item Transfer curves for BLG-on-CrBr$_3$ and BLG-on-CrI$_3$ devices; Magnetic field dependence of charge neutrality point.
\end{itemize}

\vspace{0.6cm}

\section*{Acknowledgements}
The authors gratefully acknowledge Alexandre Ferreira for technical support, and Nicolas Ubrig and Sergey Slizovskiy for fruitful discussions. We acknowledge support from the Swiss National  Science Foundation, the EU Graphene Flagship project, EPSRC CDT Graphene-NOWNANO, and EPSRC grants EP/S030719/1 and EP/V007033/1.  Z. W. acknowledges the National Natural Science Foundation of China (Grants no. 11904276) and the Fundamental Research Funds for the Central Universities. K. W. and T. T. acknowledge support from the Elemental Strategy Initiative conducted by the MEXT, Japan (Grant Number JPMXP0112101001) and JSPS KAKENHI (Grant Numbers 19H05790, 20H00354, and 21H05233).


\section*{Notes}
G.T. and D.S.D. contributed equally to this work.

The authors declare no competing financial interest.


%


\end{document}

%% file: incs.tex
\newcommand{\CrBrThree}{CrBr\textsubscript{3}}